\documentclass[aps,prl,twocolumn,superscriptaddress,float,amsmath]{revtex4-1}
\usepackage{xcolor}
\usepackage{graphicx}
\usepackage{amsmath,bm,epsfig}
\definecolor{g-blue}{rgb}{0.83,0.95,1}
\usepackage{bm}
\usepackage{amssymb}
\usepackage{amsfonts}
\usepackage{amsmath}
\usepackage{graphicx}
\usepackage{ulem}
\usepackage[pdftex,colorlinks=true,allcolors=blue,breaklinks=true]{hyperref}  

\makeatletter
\def\set@curr@file#1{%
	\begingroup
	\escapechar\m@ne
	\xdef\@curr@file{\expandafter\string\csname #1\endcsname}%
	\endgroup
}
\def\quote@name#1{"\quote@@name#1\@gobble""}
\def\quote@@name#1"{#1\quote@@name}
\def\unquote@name#1{\quote@@name#1\@gobble"}
\makeatother

\newcommand{\Eq}[1]{Eq.\,(\ref{#1})}
\newcommand{\Fig}[1]{Fig.\,\ref{#1}}
\def\Fbox#1{\vskip1ex\hbox to 8.5cm{\hfil\fboxsep0.3cm\fbox{%
			\parbox{8.0cm}{#1}}\hfil}\vskip1ex\noindent}  

\renewcommand{\sb}[1]{_{\text {#1}}}  
\def\Sb#1{_{\scriptscriptstyle\rm{#1}}}
\def\He4 {$^4$He~}

\begin{document}
	
	\title{Evolution of room-temperature magnon gas \\ toward coherent Bose-Einstein condensate}
	 
	\author{Timo~B.~Noack}
	\email{tnoack@rhrk.uni-kl.de}
	\affiliation{Fachbereich Physik and Landesforschungszentrum OPTIMAS, Technische Universität Kaiserslautern, 67663 Kaiserslautern, Germany}
	
	\author{Vitaliy~I.~Vasyuchka}
	\affiliation{Fachbereich Physik and Landesforschungszentrum OPTIMAS, Technische Universität Kaiserslautern, 67663 Kaiserslautern, Germany}
	\email{vasyuchka@physik.uni-kl.de}
	
	\author{Anna~Pomyalov}
	\affiliation{Department of Chemical and Biological Physics, Weizmann Institute of Science, Rehovot 76100, Israel}
	
	\author{Victor~S.~L’vov}
	\affiliation{Department of Chemical and Biological Physics, Weizmann Institute of Science, Rehovot 76100, Israel}
	
	\author{Alexander~A.~Serga}
	\affiliation{Fachbereich Physik and Landesforschungszentrum OPTIMAS, Technische Universität Kaiserslautern, 67663 Kaiserslautern, Germany}
	
	\author{Burkard~Hillebrands}
	\affiliation{Fachbereich Physik and Landesforschungszentrum OPTIMAS, Technische Universität Kaiserslautern, 67663 Kaiserslautern, Germany}
	
	\date{\today}
	
	\begin{abstract}
	The appearance of spontaneous coherence is a fundamental feature of a Bose-Einstein condensate and an essential requirement for possible applications of the condensates for data processing and quantum computing. In the case of a magnon condensate in a magnetic crystal, such computing can be performed even at room temperature. So far, the process of coherence formation in a magnon condensate was inaccessible. We study the evolution of magnon radiation spectra by direct detection of microwave radiation emitted by magnons in a parametrically driven yttrium iron garnet crystal. By using specially shaped bulk samples, we show that the parametrically overpopulated magnon gas evolves to a state, whose coherence is only limited by the natural magnon relaxation into the crystal lattice. 
	\end{abstract}
	
	
	\maketitle
	
	The Bose-Einstein condensate (BEC) is a state of matter encompassing a macroscopically large number of bosons that occupy the lowest quantum state, demonstrating  coherence at macroscopic scales \cite{stone2015einstein,einstein1924quantentheorie,bose1924plancks,froelich1968quasiparticles,snoke2006coherence}. This phenomenon was observed and investigated in atomic systems such as $^4$He, $^3$He (where the role of bosons is played by Cooper pairs of fermionic $^3$He atoms), and in ultra-cold trapped atoms \cite{davis1995bose,anderson1995observation}. 
	BECs were also found in systems of bosonic quasiparticles such as polaritons \cite{Amo2009polaritons} and excitons \cite{eisenstein2004excitons} in semiconductors, photons in micro-cavities \cite{klaers2010photons}, as well as magnons in superfluid $^3$He \cite{bunkov2008BECinHe} and magnetic crystals \cite{demokritov2006bose,serga2014bose,schneider2020rapid_cooling}. \looseness=-1
	
	The presence of macroscopic coherence is of fundamental importance for understanding the physical properties of BECs, including such exciting phenomena as superconductivity and superfluidity. Furthermore, there is a range of novel effects and applications that exploit the coherence of macroscopic BEC wave functions \cite{nakata2014persistent_current, tserkovnyak2017, byrnes2012, adrianov2014, Xue2019polartiton_qubit, Ghosh2020exciton-polariton_quantum-comp}, especially in the rapidly developing field of quantum computing \cite{byrnes2012,adrianov2014,Xue2019polartiton_qubit,Ghosh2020exciton-polariton_quantum-comp}. Unlike already demonstrated superconductor-based quantum computers, which operate at temperatures around 20\,$\mu$K \cite{arute2019IBM}, BEC-based qubits can be implemented at significantly higher temperatures. For instance, a magnon BEC in ferrimagnetic yttrium iron garnet (Y$_3$Fe$_5$O$_{12}$, YIG) \cite{cherepanov1993saga_YIG} crystals is formed even at room temperature \cite{bozhko2016}. \looseness=-1
	
	The magnon condensate is usually created in YIG by parametric pumping of magnons in an external microwave electromagnetic field. In this process \cite{schloemann1962, NSW-book}, external microwave photons of frequency $\omega_\mathrm{p}$ and wavenumber $q_\mathrm{p} \simeq 0$ split into two magnons with the frequency $\omega\sb{m}=\omega_\mathrm{p}/2$ and wavevectors $\pm \mathbf{q}_\mathrm{m}$. They populate a gaseous magnon distribution with internal interactions provided by the four-magnon scattering processes $2\! \Leftrightarrow \!2$. Eventually the magnon gas thermalizes to the bottom of the frequency spectrum \cite{ZLF-book} and forms a Bose-Einstein condensate there \cite{demokritov2006bose}. In in-plane magnetized YIG films, magnons condense at two equivalent frequency minima $\omega_\mathrm{min}(\mathbf{q})$ with $\mathbf{q} = \pm \mathbf{q}_{_\mathrm{BEC}}$.
	
	The magnon BEC is conveniently studied by means of Brillouin light scattering (BLS) spectroscopy \cite{demokritov2006bose,serga2014bose} delivering information about the magnon spectral density distribution. Unfortunately, due to the limited frequency resolution of the optical Fabry-P\'{e}rot interferometers used in BLS facilities, the coherence of a magnon BEC cannot be proven directly.
	Due to the phase insensitivity of the Brillouin light scattering process, studies of the BEC relaxation dynamics employing time-resolved BLS spectroscopy fail to account for BEC dephasing. The insufficient frequency resolution makes it impossible to separate the relaxation dynamics of condensed and thermal magnons. Moreover, the possible outflow of the condensate from a spatially localized probing light spot complicates the interpretation of the obtained experimental results (see \cite{demidov2008} and the corresponding discussion in \cite{kreil2019time_crystal}).
	
	Alternatively, magnon BEC coherence can be tested indirectly by observation of phenomena such as quantized vorticity \cite{Nowik-Boltyk2012}, supercurrents \cite{bozhko2016supercurrent}, Bogoliubov waves \cite{bozhko2019Bogoliubov_waves}, or Josephson oscillations \cite{kreil2019josephson}, which are canonical features of both atomic and quasiparticle quantum condensates. 
	Our studies of some of these phenomena \cite{bozhko2016supercurrent, bozhko2019Bogoliubov_waves, kreil2019josephson, kreil2018from_KI_to_supercurrents, mihalceanu2019low_temp_BEC} have shown that they occur only in a freely evolving magnon gas after switching off the microwave pumping. This takes place probably because the intense pumping process prevents condensation by heating the magnon gas \cite{serga2014bose} and mixing the magnon frequencies near the bottom of their spectra \cite{kreil2019time_crystal}. 
	The observation of these effects indicates the presence of a time-dependent BEC coherence, but leaves open the question about the degree of coherence.
	
 	Attempts to qualitatively characterize BEC coherence were made using a novel high-resolution magneto-optical Kerr-effect spectroscopy \cite{dzyapko2016, dzyapko2017},  microwave spectroscopy of electromagnetic signals emitted at the ferromagnetic resonance frequency due to the confluence of bottom magnons with opposite wavevectors \cite{dzyapko2008, rezende2009}, and by BLS observations of the interference of $\pm \mathbf{q}_{_\mathrm{BEC}}$ magnon condensates \cite{Nowik-Boltyk2012}. 
	They demonstrate a very low modulation depth of the interference pattern \cite{Nowik-Boltyk2012}, 
 	a rather broad frequency spectral BEC line \cite{dzyapko2008}, and
 	increase in the BEC line width when the pumping power exceeds the threshold of BEC formation \cite{dzyapko2017}. 
 	These results themselves are certainly interesting and important. However, without additional data on the temporal evolution of coherence, their interpretation is difficult and remains questionable.

\begin{figure}  
		\includegraphics[width=1.0\linewidth]{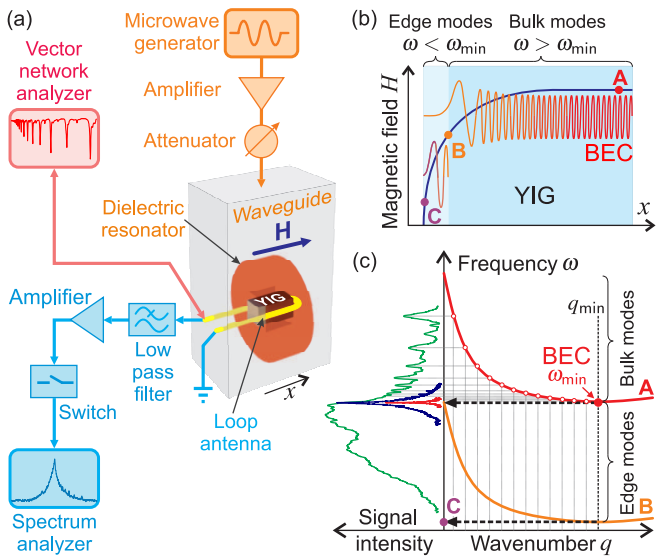}
		\caption{(a) The experimental setup for microwave detection of magnon dynamics. Various parts of the setup are color-coded: orange denotes the pumping circuit, blue highlights the receiving circuit, and red marks the test circuit. (b) Schematic representation of the bulk BEC mode and one of the edge magnon modes in a cuboid YIG sample. The monotonic blue line shows the profile of the static magnetic field $\mathbf{H}$ within the YIG sample. Color points denote three field values: A -- deeply inside the sample, B -- at the point near the sample edge, where the bulk BEC mode becomes evanescent with purely imaginary wavenumber, C -- at the sample edge. (c) Schematic representation of magnon dispersion curves in the middle of the sample (at point A) and near the edge (at point B). The green, blue, and red signal intensity lines represent the microwave power spectra from the YIG sample registered during the one-microsecond interval before the end of the pumping action, and 2\,$\mu$s and 4\,$\mu$s after the pump pulse is turned off, respectively.
		}
		\label{F:Setup}
\end{figure}
	
\begin{figure*}  
		\includegraphics[width=1.0\linewidth]{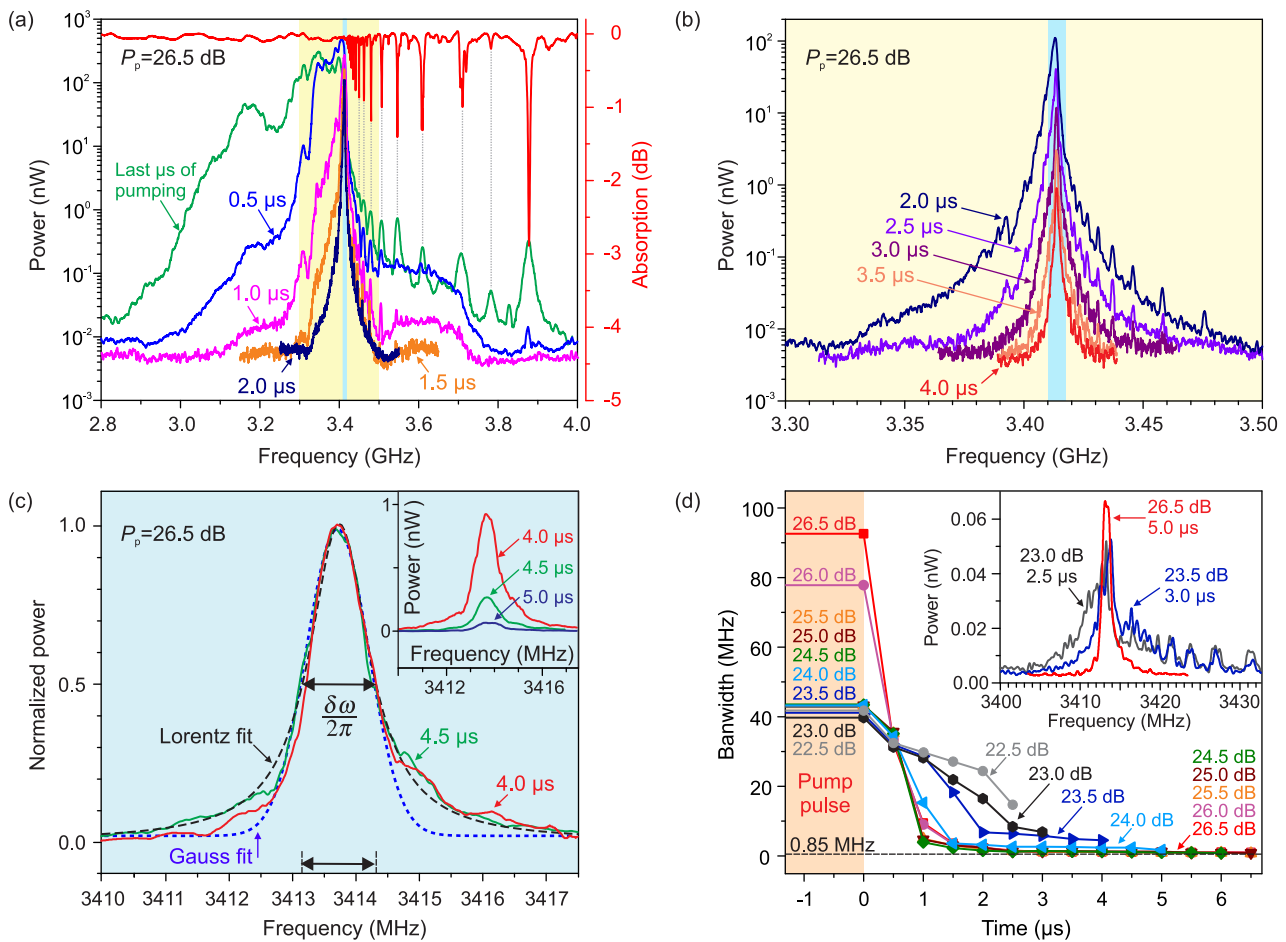}
		\caption{Magnon radiation spectra $J\sb{rad}(\omega,t)$ for the pumping power $P\sb p=26.5$\,dB above the parametric instability threshold and $\mu_0 H = 115$\,mT 
		(a), (b) and (c):  The spectra are takenat consequent moments of time before [green line in panel (a)] and after switching off the pumping power. The yellow and blue shading in (a)-(c) denote the frequency range: yellow shading corresponds to the spectra shown in (b), blue shading corresponds to the spectra shown in (c). The absorption spectrum $J\sb{abs}(\omega,t)$ measured without pumping is shown in (a) by the red line.
		In panel (c), the spectra are normalized by their maximum value. A Lorentz fit is shown by the black dashed line, the Gaussian fit is indicated by the blue dotted line. The inset shows the same spectra without normalization. (d): Time evolution of the signal bandwidth for different pumping powers $P\sb p$. The black dashed line in (d) marks the linewidth $\delta \omega_\mathrm{fin}/(2\pi) = 0.87$ MHz. The inset demonstrates the narrowing of the radiation spectrum of the freely evolving magnon gas with increase of the pumping power $P\sb p$. The different delay times are chosen to have approximately the same maximum magnitude for the three spectra shown.  
				}
		\label{F:Spectra}
\end{figure*}
		
The main goal of this work is to understand the time evolution of the magnon gas toward a coherent BEC state. By direct measurement of microwave radiation from a bulk YIG sample, we show that the frequency-broadband emission spectrum, detected during the pumping action, transforms after the end of pumping into a sharp spectral peak at the lowest frequency of the magnon spectrum. This peak is earlier formed and gets pronounced with increasing pumping power and, consequently, with the density of parametrically pumped magnons. At high pumping powers, the peak has a Lorentz shape and its width is consistent with the magnetic relaxation frequency into the YIG crystal lattice. The appearance of this peak is associated with the formation of the magnon BEC, whose coherence is, therefore, limited only by the natural magnon decay.

In YIG films used in all previous BEC studies, the condensed magnons have a wavelength of about a few micrometers and are thus weakly coupled to the electromagnetic field, making them difficult to detect via directly emitted radiation. The main idea of our experiment is to use a YIG cuboid bulk sample to enhance this coupling. The experimental setup is shown in \Fig{F:Setup}(a). The YIG sample sized $0.3\times0.3\times0.6\,$mm$^3$ is magnetized along its long side, which is oriented along the $x$ coordinate axis. 
Due to the demagnetization effect, the static magnetic field $H(x)$ inside such a sample [blue line in \Fig{F:Setup}(b)] is smaller at its edges than in the middle. For a slowly spatially-varying magnetic field $H(x)$, the magnon frequency may be considered as an adiabatic invariant: $\omega\big[\mathbf{q}(x),H(x)\big]=\mathrm{const}$, while the wavevector becomes position-dependent $\mathbf{q} \to \mathbf{q} (x)$ \cite{schloemann1964, smith2008spin}. 

For the BEC magnons, this frequency is equal to the frequency of the spectrum minimum $\omega\sb{min}$ in the central part of the sample [marked by point A in \Fig{F:Setup}(b)]:   
	\begin{equation}\label{qx}
		\omega\big[q(x),H(x)\big]=	\omega\sb{min}=\omega(q\sb{min},H\Sb A ) \,,
	\end{equation} 
providing a relation between $q(x)$ and $H(x)$. 

The bulk frequency spectrum $\omega(q,H\Sb A)$ is schematically shown by the red line in the upper part of \Fig{F:Setup}(c). As one moves from point A to some point B near the sample edge, the magnetic field decreases and the spectrum branch $\omega\big[q(x),H(x)\big]$ is continuously shifted down. The spectrum $\omega(q,H\Sb B)$ for the lower magnetic field at point B is schematically shown by the orange line in the lower part of \Fig{F:Setup}(c). Therefore, according to \Eq{qx}, the wavenumber $q(x)$ of the BEC magnons with $\omega \big[q(x),H(x)\big]= \omega\sb{min}$ decreases towards the edges of the sample, reaching zero value for $x=x\Sb B$ as is indicated by the black dashed arrow in \Fig{F:Setup}(c). For $x<x\Sb B$, the bulk mode becomes evanescent with a purely imaginary wavenumber. In the near-edge region, between points B and C, only localized edge modes exist. A small value of $q(x)$ near point B, and, correspondingly, a large wavelength of magnons, enhances the coupling of the magnon BEC with the electromagnetic field.  

The large volume of the sample and its cuboidal shape make it possible to achieve the desired detection sensitivity using a simple inductive loop antenna placed around the sample and connected to the receiving circuit marked in blue in \Fig{F:Setup}(a). The fast microwave switch is used to measure power-frequency radiation spectra $J\sb{rad}(\omega,t)$ in $1\,\mu$s-long time windows shifted by $0.5\,\mu$s steps. The low-pass filter protects the spectrum analyzer from a strong pumping signal. Magnons are pumped by 6\,$\mu$s-long pulses of the electromagnetic field of frequency $\omega\sb p = 2\pi \cdot 7.68\,$GHz, whose amplitude is enhanced by a dielectric resonator (see \Fig{F:Setup}(a), where the orange circuitry illustrates the pumping circuit). 

Consider first the structure of the eigenmodes of the cuboid sample. Their absorption spectrum $J\sb{abs}(\omega,t)$, measured by a vector network analyzer and colored in red in \Fig{F:Setup}(a), is shown by the red line in \Fig{F:Spectra}(a). In the same figure, the green line denotes the radiation spectrum $J\sb{rad}(\omega,t)$ of the sample measured during the last microsecond of pumping.
Above $\omega/(2\pi) > 3.41\,$GHz, one can see a set of discrete peaks, whose frequencies coincide \cite{14MHz} in both spectra [see thin vertical dashed lines in \Fig{F:Setup}(a)]. 
They originate from the bulk magnon modes, schematically shown on the magnon dispersion branch A in \Fig{F:Setup}(c). In an infinite sample, the spectrum of such modes is continuous. However, in the finite sample, only a discrete set of  wavenumbers $q_n$ is allowed. In a simple case of a longitudinally magnetized bar of length $L$, the periodic boundary conditions dictate $q_n=2\pi n/L$. They are illustrated in \Fig{F:Setup}(c) by gray dotted vertical lines. The  corresponding ``allowed" values of $\omega=\omega_n=\omega(q_n)$ are shown by empty dots and horizontal gray lines. Larger values of $\omega(q_n)$ correspond to smaller $q_n$, which are better coupled with the inductive loop. This explains why the peaks at higher frequencies are more pronounced in \Fig{F:Spectra}(a). Furthermore, the peak positions become closer as $\omega$ approaches $\omega\sb{min}$ from above. This behavior is well reproduced by the spectra in \Fig{F:Spectra}(a), where $\omega\sb{min}/(2\pi) = 3.41\,$GHz. 
	
The part of the spectra at $\omega < \omega\sb{min}$ originates from the modes localized near the sample edges. Indeed, the decreasing intrinsic magnetic field [blue line in \Fig{F:Setup}(b)] between B and the edge of the sample serves as a potential well. In this well, there exists a discrete set of magnon states having a relatively large characteristic scale. These edge modes are well coupled with the electromagnetic field around the sample and therefore are affected by  additional radiation damping. Since the additional damping results in a low quality factor of these modes, their discrete structure is hardly visible in the radiation spectrum. For the same reason, these modes practically do not contribute to the absorption spectrum. Note also that the actual positions of the peaks  in \Fig{F:Spectra}(a) are not so regular as expected from the one-dimensional model. In a finite sample of a general shape, the role of ``allowed" $\omega_n$ is played by the frequencies of so-called Walker modes in a cuboid, which may be not equidistant \cite{Melkov-book}. 

Consider now the evolution of the radiation spectrum $J\sb{rad}(\omega,t)$. During the pumping, it extends from 2.8\,GHz to 4\,GHz as is indicated by the green line in \Fig{F:Spectra}(a). The main radiation power is located in the 100\,MHz band around $\omega\sb{min}$. Such a large width is caused by intensive shaking of the entire magnon frequency spectra by a powerful microwave pumping field. For instance, for $P_\mathrm{p}=26.5$\,dB, the amplitude of the microwave pumping field $h_\mathrm{p}$ applied parallel to the bias magnetic field $H$ is estimated to be about 25\,Oe. As a result, the magnon frequency spectrum moves up and down in the range of $\pm 70$\,MHz, which is close to the radiation spectrum width.

After switching off the pumping power, the shaking of the magnon frequencies ceases and the spectrum width quickly decreases as seen in \Fig{F:Spectra}(a-c). The edge modes with $\omega<\omega\sb{min}$ uniformly decay within the first 2\,$\mu$s, likely due to effective radiation damping. The evolution of the bulk modes with $\omega\gtrsim \omega\sb{min}$ is more complicated. The most intense peaks in the initial spectrum are strongly decreased already within a time interval of 0.5\,$\mu$s, especially at  frequencies, for which the radiation damping is most efficient. Another reason for the spectrum narrowing is the redistribution of magnons towards modes with $\omega\simeq \omega\sb{min}$ during the BEC formation.

In \Fig{F:Spectra}(b) and (c), we show details of the further evolution of $J\sb{rad}(\omega,t)$. Here we plot the spectra for more narrow frequency intervals, colored in \Fig{F:Spectra}(a) by yellow and blue shading, respectively. The width of the spectra measured with time delays $t_\mathrm{d} =$ (2--4)\,$\mu$s decreases further, see \Fig{F:Spectra}(b), confirming the process of gathering the magnons toward the BEC state at $\omega =\omega\sb{min}$. At later times, the width and form of the spectra practically do not change. In \Fig{F:Spectra}(c), we show two spectra, for $t_\mathrm{d}=4.0\,\mu$s and $4.5\,\mu$s, normalized by their maximum value. The normalized spectra almost coincide, while their magnitude continues to decrease in time [see inset in \Fig{F:Spectra}(c)]. \looseness=-1
 	
To quantify the radiation spectra, we investigate their bandwidth $\delta \omega$. For  single-peak spectra, we chose $\delta \omega$ as the peak width at the half-maximum magnitude. This definition corresponds to the width of the Lorentz peak, describing a uniformly broadened spectral line. For the spectra with complex many-peak structure, such as the spectra in \Fig{F:Spectra}(a), we generalize this definition as follows:
	\begin{eqnarray}\label{def-om}
 	\delta \omega =    \sqrt{2 \int \Omega^2 \widetilde{f} (\Omega) d\Omega \, \Big/ \int  \widetilde{f}(\Omega) d\Omega }  \,,
	\end{eqnarray} 
where $\Omega= \omega- \omega\sb{min}$ and $\widetilde f(\Omega)$ is the truncated spectrum $f(\Omega)$, with the spectrum part below 5.5\% of its maximum magnitude removed.
The time evolution of the bandwidth $\delta\omega(t)$ for different $P\sb p$ from 22.5\,dB to 26.5\,dB above the threshold of the parametric instability, is shown in \Fig{F:Spectra}(d).
The bandwidth during the pump pulse ($t<0$) is larger for larger $P\sb p$. After the pumping is turned off, $\delta \omega$ decreases monotonically due to the Bose-Einstein condensation process. This process is dominated by four-magnon scattering processes with a rate proportional to $N^2$ \cite{ZLF-book, NSW-book}, where $N$ is the number of bottom magnons. Increasing $N$ at larger $P\sb p$ leads to more efficient magnon gathering toward $\omega\sb{min}$ and a faster decrease in $\delta \omega$. This narrowing has a threshold character and occurs when the pumping power increases from 22.5\,dB to 24\,dB. We consider this as additional evidence of magnon condensate formation at $P\sb p \gtrsim 24$\,dB.

The insert in \Fig{F:Spectra}(d) presents $J\sb{rad}(\omega,t)$ spectra measured near the detection limit of the experimental setup for two low pumping powers of $P\sb p=23.0$\,dB and 23.5\,dB, and for the highest value  of $P\sb p=26.5\,$dB. Being rather weak, they correspond to the final stages of the evolution of the magnon system at the bottom of their spectrum, when no non-linear scattering is expected and both condensed and gaseous magnons linearly decay to the thermal phonon bath. However, the structure of these residual spectra is determined by the previous processes of nonlinear four-magnon scattering and BEC formation. For weaker pumping, the spectral line at $\omega\sb{min}$ is surrounded by a distribution of relatively strongly populated magnon modes, which demonstrate a clear comb-like structure at frequencies above $\omega\sb{min}$. Increasing pumping power leads to the de-population of all these modes due to magnon gathering toward the dense BEC. As a result, only the spectral line related to the magnon condensate remains in the spectrum.

At high $P\sb p$, the residual spectra are best fitted with the Lorentz function
\begin{align}\label{Lor} 
J\sb{rad} (\omega,t) = \frac{  I\sb{rad} (t)\,\delta\omega  }{(\omega - \omega\sb{min})^2+ \delta\omega  ^2/  4 } \,,  
\end{align}
in which $\delta\omega $ is the bandwidth of the frequency spectra and $I\sb{rad} (t)$ is the time-dependent total power of the signal. The fit is shown in \Fig{F:Spectra}(c).
Another possible (Gaussian) shape  is indicated in \Fig{F:Spectra}(c) by the  blue dotted line for comparison. 

Probably the most important evidence for coherency, as shown in \Fig{F:Spectra}, is that at later times (say, after the time delay $t_\mathrm{d}>2.5\,\mu$s) the exponentially decaying residual spectra for $P\sb p \geq24\,$dB have a near-Lorentzian shape\,\eqref{Lor} with the bandwidth $\delta\omega$ approaching the value of about $\delta\omega\sb{fin}/(2\pi)\simeq$ 0.85\,MHz, almost independent of $P\sb p$. \looseness=-1
	 
To summarize, the magnon system evolving toward BEC reaches full coherence, with the width of the magnon radiation spectrum decreasing by more than two orders of magnitude. The residual bandwidth is mainly determined by the lifetime of magnons, as expected for a fully coherent BEC consisting of a single magnon state. Moreover, we show that a coupling of the magnon BEC with dynamic stray fields outside the sample is enabled by a proper choice of the sample shape giving direct spectroscopic access to the BEC. Such an approach can function as a convenient tool for integrating magnetic quantum systems into electrical environments. We believe that this direct demonstration of the magnon BEC coherence brings closer the implementation of room temperature BEC-based computing. \looseness=-1

\begin{acknowledgments} 
	This research was funded by the European Research Council within the Advanced Grant No. 694709 “SuperMagno\-nics” and by the Deutsche Forschungs\-gemein\-schaft (DFG, German Research Foundation) within the Transregional Collaborative Research Center -- TRR 173 -- 268565370 ``Spin+X'' (project B01). The authors are grateful to G.\,A.\,Melkov and H.\,Yu.\,Musiienko-Shmarova for fruitful discussions.
\end{acknowledgments}


\begin{thebibliography}{99}
\bibitem{14MHz}
	In fact, the frequencies of the peaks of the radiation spectrum measured during pump action are 14\,MHz higher than the frequencies of the corresponding peaks in all other radiation and absorption spectra. This difference arises due to decrease in the sample magnetization caused by pumping magnons. To demonstrate the identity of the structure of all spectra, we shifted this radiation spectrum [the green line in Fig.\,\ref{F:Spectra}(a)] down by 14\,MHz. \
\end{thebibliography}

\end{document}